\documentclass[runningheads]{svmult}

\usepackage{makeidx}   
\usepackage{graphicx}  
\usepackage{subeqnar}  
\usepackage{multicol}  
\usepackage{physprbb}  
\makeindex             

\newcommand{\degr}{\hbox{$^\circ$}}
\newcommand{\arcmin}{\hbox{$^\prime$}}

\newcommand{\utw}{\smash{\rlap{\lower5pt\hbox{$\sim$}}}}
\newcommand{\udtw}{\smash{\rlap{\lower6pt\hbox{$\approx$}}}}

\newcommand{\farcm}{\hbox{$.\mkern-4mu^\prime$}}
\newcommand{\farcs}{\hbox{$.\!\!^{\prime\prime}$}}

\begin{document}
\title*{The PNLF distance to the Sculptor Group galaxy NGC 55}

\titlerunning{PNLF distance to NGC 55}

\author{G. C. Van de Steene\inst{1}
\and G. H. Jacoby\inst{2}
\and C. Praet\inst{3}
\and R. Ciardullo\inst{4}
\and H. Dejonghe\inst{3} }

\authorrunning{G. C. Van de Steene et al.}

\institute{Royal Observatory of Belgium, Ringlaan 3, 1180 Brussels, Belgium
\and WIYN Observatory, PO Box 26732, Tucson, AZ, 85726, USA
\and Universiteit Gent, Krijgslaan 281 (S9), 9000 Gent, Belgium
\and Department of Astronomy and Astrophysics, Penn State University, 525 Davey Lab University Park, PA 16802, USA }

\maketitle

\begin{abstract}

We identified 21 new Planetary Nebula (PN) candidates in the Sculptor
Group galaxy NGC~55. We determined a most likely distance of
2.00~$\pm$~0.2~Mpc using the Planetary Nebulae Luminosity Function (PNLF)
method.  The distance to NGC~55 is larger than previously
determined distances, which means that the Sculptor Group is a bit further
away from the Local Group than previously thought. The distance to
NGC~55 is again similar to the distance of NGC~300, adding support to
the suggestion that these galaxies form a bound pair.

\end{abstract}

\section{NGC 55}

NGC~55 is a highly inclined (i$=$$85\pm5\degr$) late type galaxy
(SB(s)m) in the Sculptor Group. Its Holmberg radius is 20\farcm2 and
blue magnitude m$_B$$=$8.42.  Sculptor is a very loosely concentrated
and almost freely expanding aggregate of galaxies of prolate shape
which we view pole-on.  It stretches from the outskirts of the Local
Group at 1.5 Mpc out to 5 Mpc. Besides the 5 main spiral galaxies, the
Sculptor Group contains several dwarf galaxies (Jerjen et
al. \cite{Jerjen98}).

NGC~55 is one of the 5 bright spiral galaxies of the Sculptor Group,
which could be the one nearest to the Milky Way.  Distance estimates
for NGC~55 range from 1.34~Mpc based on carbon stars (Pritchet et
al. \cite{Pritchet87}) to 1.8~Mpc based on the Tully-Fisher relation
(Karachentsev et al. \cite{Karachentsev03}). The two bright spiral
galaxies NGC~55 and NGC~300 are both at the near side of the Sculptor
group and it has been suggested that NGC~55 and NGC~300 form a bound
pair (Graham \cite{Graham82}; Pritchet et al.  \cite{Pritchet87};
Whiting \cite{Whiting99}). The distance to NGC~300 is well determined
via Cepheids (Freedman et al. \cite{Freedman01}: 2.07~$\pm$~0.07~Mpc)
and via the PLNF method (Soffner et al. \cite{Soffner96}: 2.4 $\pm$
0.4 Mpc). NGC~55 has remained without a reliable distance estimate.

\section{Observations}        

The observations were done with the Wide Field Imager (WFI) on the
2.2-m telescope at the La Silla Observatory of the European Southern
Observatory on the 4 and 5 July 1999. The weather was photometric.
WFI consists of a 4 times 2 mosaic of 2k~x~4k CCDs. The pixel scale is
0\farcs238 and the total field of view is 34\arcmin\ x 33\arcmin,
which covers the entire galaxy.  NGC~55 was observed through the
[O~III]/8 (FWHM 80.34 \AA) and the off-band filter 518/16, and the
H$\alpha$/7 filter for a total of 3120~sec, 1620~sec, and 3600~sec
respectively. On the first day the galaxy was offset to the east
and south by 3\arcmin\ to fill in the CCD gaps. The seeing was
1\farcs4.

\section{Data reduction and results}

The {\sc esowfi} external package in {\sc iraf} was used to
convert the ESO headers to suit the {\sc mscred} package and
to set the instrument files and astrometry solution.  Next the
reductions were done using the {\sc mscred} mosaic reduction package
in {\sc iraf} according to the guide by Valdes (\cite{Valdes98}).

We followed the survey technique as described in e.g. Jacoby et
al.\cite{Jacoby89b} and Ciardullo et al. \cite{Ciardullo89a}.  The on-
and off-band images were aligned.  The off-band images were then
scaled to the level of the on-band images and subtracted from them to
produce the difference image.  We identified the PNe by blinking this
difference image with the off-band image, and the H$\alpha$+[N II]
images. In order to discriminate PNe from other emission-line sources,
we used the following critieria (Feldmeier et al. \cite{Feldmeier97}):
(1) PN candidates had to have a point-spread function (PSF) consistent
with that of a point source as all PNe are expected to be unresolved
at the distance of NGC~55; (2) PN candidates had to be invisible on
the off-band image to exclude bright OB stars exciting H~II regions;
and (3) PN candidates had to be significantly fainter in H$\alpha$+[N
II] image than [O III], in order to reduce further the possibility of
contamination from H~II regions.  With all these constraints, only
compact, H~II regions that have high nebula excitation and faint
central OB associations will have been mistaken for PNe.  However,
even these will be faint and hence cannot significantly affect the PNLF.
In total we identified 21 candidate planetaries in NGC~55.

The PN candidates were measured photometrically in the off-band image
using {\sc phot} in {\sc iraf} and flux calibrated using standard
stars and the procedures outlined in Jacoby et al. \cite{Jacoby87}.
In order to determine the filter transmission for the PNe we need to
take the redshift into account and the bandpass shift of
the interference filter to the blue.  The systemic velocity of NGC~55
was taken to be 116~km/s (Puche \& Carignan \cite{Puche91}).  
The percentage of transmission is about
88\% across the filter for the [O~III] line and about 87\% for the
H$\alpha$ filter at the location of our objects.  The difference in
transmission across the field has negligible effect on the derived
flux values and magnitudes.

The resulting monochromatic [O~III] flux values F$_{5007}$ were
converted to [O~III]$\lambda$5007 magnitudes using (Jacoby \cite{Jacoby89a}) : 

\begin{equation}
m_{5007}  =  -2.5~log(F_{5007}) - 13.74 .
\end{equation}

\section{The PNLF distance} 

[O~III] magnitudes were corrected for the interstellar extinction.  We
consider only the foreground Galactic extinction towards NGC~55 and
adopt E(B-V)$=$0.013 mag (Schlegel et al. \cite{Schlegel98}).

Identifications become incomplete beyond m$_{5007}$~$=$~23.5~mag. 
Besides missing faint PNe, the probability of overlap with a star or
with an H~II region increases towards the centre, particularly in
the H$\alpha$+[N~II] image.  Hence brighter PNe than the ones
recovered in the outskirts may have been missed towards the center.

\begin{figure}
\includegraphics[width=.95\textwidth]{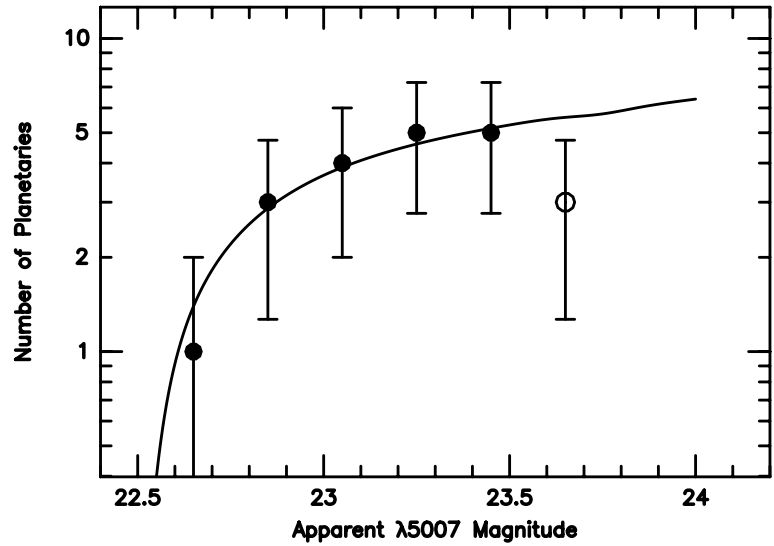}
\caption{Observed [O III]$\lambda$5007 PNLF of NGC~55. The curve represents
the best-fitting empirical PNLF convolved with the photometric error function 
and shifted to the most likely distance. The open circle represents a
point past the completeness limit.}
\label{vandesteeneF1} 
\end{figure}

From these data the PNLF distance to the galaxy can normally be derived by
convolving the empirical model for the PNLF given by Ciardullo et al. 
\cite{Ciardullo89b}:
\begin{equation}
N(m) \propto  e^{0.307M} [1-e^{3(M^*-M)}]
\end{equation}
with the photometric error function and fitting the data to the
resultant curve via the method of maximum likelihood. This takes into
account that the probability of observing PNe near the cutoff
magnitude M$^*$ decreases for small sample
sizes. Fig. \ref{vandesteeneF1} plots the PNLF for the
sample. Assuming M$^*$$=$~$-$4.47, based on the calibration to M~31
(Ciardullo et al. \cite{Ciardullo02}), we obtain a most likely
distance modulus (m~$-$~M$^*$)~$=$~26.95.  The PNLF cut-off is fainter
in small, low metallicity galaxies, but well modeled by the
theoretical relation of Dopita et al. \cite{Dopita92}.
The metallicity correction as determined from the oxygen abundance,
is needed only for galaxies with metallicities smaller than the 
LMC (12$+$log(O/H)~$<$~8.5) (Ciardullo et
al. \cite{Ciardullo02}). The oxygen abundance of NGC~55 as determined
based on H~II regions is  12$+$log(O/H)$=$8.05 
(T\"ullmann et al. \cite{Tullmann03}), which  is much
lower than this value and simlar to the SMC metallicity. Assuming a solar
abundance of oxygen of 12$+$log(O/H)~$=$~8.87 (Grevesse et
al. \cite{Grevesse96}), the metallicity corrected value of the 
distance modulus is (m~$-$~M$^*$)~$=$~26.50~$\pm$~0.2, which corresponds
to a distance of 2.00~$\pm$~0.2~Mpc.  This value is a bit larger than
previously determined distances to NGC~55, which would mean that the
Sculptor Group would be a bit further away than previously thought.

NGC~300 also has a low metallicity, though not as low as NGC~55.  The
metallicity correction to be applied is 0.15 mag (Ciardullo et
al. \cite{Ciardullo02}). The distance modulus to NGC~300 determined
via the PNLF maximum likelihood method is 26.8 mag (Soffner et
al.~\cite{Soffner96}) and after correction for metallicity 26.65 mag.
This corresponds to a distance of 2.14~$\pm$0.4 Mpc , which is in very
good agreement with the Cepheid distance determination of
2.02~$\pm$~0.07 Mpc by Freedman et al. (\cite{Freedman01}).

The PNLF distances to NGC~55 and NGC~300 are again similar, which adds
support to the fact that they form a bound pair and illustrates the
consistency of the PNLF method for distance determination.

\section{Conclusions}

We identified 21 new PNe candidates in the Sculptor Group galaxy
NGC~55.  The PNLF method gives us a most likely distance of
2.00~$\pm$0.2~Mpc, which would mean that the Sculptor Group is a bit
further away from the Local Group than previously thought.  The
distance to NGC~55 turns out to be similar to the distance of NGC~300,
adding support to the suggestion that these galaxies form a bound pair
and illustrates the consistency of the PNLF method for distance
determination.

\end{document}